\documentclass[12pt,iop,preprint, tighten]{aastex}
\usepackage{multirow,color}
\usepackage{bbding}
\usepackage{amssymb}
\usepackage{longtable}
\usepackage{array}
\usepackage{multicol}
\usepackage{lipsum}
\usepackage{color}

\usepackage[]{times, graphicx}
\def\gsim{\;\lower4pt\hbox{${\buildrel\displaystyle >\over\sim}$}\;}
\def\lsim{\;\lower4pt\hbox{${\buildrel\displaystyle <\over\sim}$}\;}
\def\grls{\;\lower4pt\hbox{${\buildrel\displaystyle >\over <}$}\;}

\def\A{{Alfv\'en}}

\def\Aic{Alfv\'enic}
\def\Aiic{Alfv\'en(ic)}

\def\ma{magnetoacoustic }

\begin{document}

\shorttitle{Turbulence Onset}
\shortauthors{J. Liu et~al.}
\title{Statistical Evidence for the Existence of {\Aic} Turbulence in Solar Coronal Loops}

\author{Jiajia Liu\altaffilmark{1,2}, Scott W. McIntosh\altaffilmark{2}, Ineke De Moortel\altaffilmark{3}, James Threlfall\altaffilmark{3}, Christian Bethge\altaffilmark{2,4}} 
\altaffiltext{1}{Earth and Space Science School, University of Science and Technology of China, NO. 96, JinZhai Road, Hefei, China}
\altaffiltext{2}{High Altitude Observatory, National Center for Atmospheric Research, P.O. Box 3000, Boulder, CO 80307, USA.}
\altaffiltext{3}{School of Mathematics and Statistics, University of St Andrews, St Andrews, Fife, KY16 9SS, UK.}
\altaffiltext{4}{Kiepenheuer Institute for Solar Physics, Freiburg, Germany}

\begin{abstract}
Recent observations have demonstrated that waves which are capable of carrying large amounts of energy are ubiquitous throughout the solar corona. However, the question of how this wave energy is dissipated (on which time and length scales) and released into the plasma remains largely unanswered. Both analytic and numerical models have previously shown that {\Aic} turbulence may play a key role not only in the generation of the fast solar wind, but in the heating of coronal loops. In an effort to bridge the gap between theory and observations, we expand on a recent study [De Moortel et al., ApJL, 782:L34, 2014] by analyzing thirty-seven clearly isolated coronal loops using data from the Coronal Multi-channel Polarimeter (CoMP) instrument. We observe {\Aic} perturbations with phase speeds which range from $250-750 \,\rm{km\,s}^{-1}$ and periods from $140-270 \,\rm{s}$ for the chosen loops. While excesses of high frequency wave-power are observed near the apex of some loops (tentatively supporting the onset of {\Aic} turbulence), we show that this excess depends on loop length and the wavelength of the observed oscillations. In deriving a proportional relationship between the loop length/wavelength ratio and the enhanced wave power at the loop apex, and from the analysis of the line-widths associated with these loops, our findings are supportive of the existence of {\Aic} turbulence in coronal loops.
\end{abstract}

\keywords{Sun: corona --- waves}

\section{Introduction}
Recent advances in the field of ground- and space-based observations of the solar corona have revealed the prevalence of oscillatory/wave-like phenomena across a wide range of structures pervading the solar atmosphere \citep[e.g.,][]{DeMoortel_2012}. The growing evidence of wave propagation (and dissipation) has lead to a resurgence of interest in waves and their contribution to both the heating of the solar atmosphere and the generation of the fast solar wind \citep[e.g.,][]{Tomczyk_2007,McIntosh_2011,Parnell_2012}. Since acoustic wave-heating is thought to be efficient only over relatively low heights \citep{Osterbrock_1961}, {\ma}(MA) waves and/or {\Aiic} waves have instead been regarded as likely candidates for oscillatory/wave-like phenomena observed along coronal structures \citep[see e.g.][]{Parker_1991, Suzuki_2005, Nakariakov_2005, McIntosh_2011, McIntosh_2012,DeMoortel_2012}.

Both fast-MA kink waves and/or {\A} waves may generate observational signatures which are typically seen as propagating transverse oscillatory displacements of the local parent structure. Such signatures have been observed along many different types of structures, including (but not limited to) coronal loops \citep[e.g.][]{Ulmschneider_1990, Aschwanden_1999, Tomczyk_2007, Threlfall_2013, Morton_2013}, jets \citep[][]{Cirtain_2007}, spicules \citep[][]{DePontieu_2007, He_2009}, prominences \citep[e.g.][]{Okamoto_2007} and coronal rain \citep[e.g.][]{Antolin_2011}. However, interpretation of these and other observations in terms of a pure fast-MA kink or {\A} wave mode alone requires additional information about the local geometry/environment of the parent structure; a substantial (and ongoing) debate regarding the underlying nature of these observations has arisen \citep[see, e.g.,][]{Erdelyi_2007, Van_Doorsselaere_2008, DeMoortel_2012}. 

Recent numerical simulations have shown that generic transverse footpoint displacements generate propagating oscillatory displacements which are composed of coupled kink-{\A} modes (e.g.~\citealt{Hood_2013, Pascoe_2013, Terradas_2010} and references therein); the term ``{\Aic}" has come to describe this inherently coupled wave mode, which cannot be entirely described by a single wave mode alone \citep[e.g.][]{McIntosh_2011}. Estimates of the energy flux carried by these waves may be sufficient to balance the energy budget of (for example) the quiet Sun \citep[e.g.][]{DeMoortel_Pascoe_2012, McIntrosh_DePontieu_2012,Goossens_2013}, and hence could form a crucial part of the energy transport into the corona and solar wind. While these waves are certainly capable of carrying the energy over large distances within the corona, one vital piece of the puzzle remains elusive; the mechanism by which energy is extracted from the waves.

Damping mechanisms are often underpinned by an energy cascade from large to small scales. The {\Aic} wave mode allows access to several such processes. For example, numerical simulations which highlight wave mode-coupling \citep[][]{Pascoe_2013} show that energy is transferred from footprint-driven (bulk) transverse motion to an azimuthal component, which can then phase-mix within a relatively small boundary layer of strong inhomogeneity. While this does lead to wave damping, it does so only in narrow regions which are (at present) beyond the resolution limit of current observations (however, it is possible that future high spatial and temporal resolution instruments, for example those on the Daniel K Inouye Solar Telescope (DKIST), may be able to shed some light on such layers, and their role in the energy transport). Transferring energy to smaller scales could lead to {\Aic} turbulence, and has been shown to be able to account for the acceleration of the fast solar wind \citep[e.g.,][]{Parker_1991, Oughton_2001, Cranmer_2005, Verdini_2010} and heating of coronal loops \citep[e.g.,][]{vanBallegooijen_2011, Asgari-Targhi_2012}. While a great deal of progress on this issue has been made using both analytical and numerical models (particularly with regard to solar wind acceleration), it is also important to relate these models to observations. The recent work by \cite{DeMoortel_2014}, which motivated this paper, has shown the tentative evidence for the onset of {\Aic} turbulence in a trans-equatorial coronal loop, by exploring the novel ``excess of high-frequency FFT (Fast Fourier Transform) power" (hereafter referred as the ``EHFF") phenomenon near the loop apex.

Following \cite{DeMoortel_2014}, we present a detailed statistical analysis of thirty-seven clearly isolated coronal loops observed in the field-of-view (FOV) of the Coronal Multichannel Polarimeter (CoMP) instrument \citep[][]{Tomczyk_2008}. This analysis suggests that the extra high-frequency part of the power spectrum (previously described as tentative evidence for the onset of {\Aic} turbulence) is particularly prevalent for coronal loops with length 3.0 times longer than the characteristic wavelength of the propagating waves. Our paper is organized as follows; in Section~\ref{sect:instr} we describe the instrument and data used in the analysis, while a detailed study of three coronal loops is presented as an example in Section~\ref{sect:cavity}, followed by a statistical survey of all our chosen loops in Section~\ref{sect:stats}. A discussion of our results can be found in Section~\ref{sect:disc} before presenting our conclusions in Section~\ref{sect:conc}.

\section{Instrument and Data}\label{sect:instr}

The Coronal Multi-channel Polarimeter \citep[CoMP;][]{Tomczyk_2008} is a combination polarimeter and narrowband tunable filter that can measure the complete polarization state in the vicinity of the 10747 \AA{} and 10798 \AA{} \ion{Fe}{13} coronal emission lines. It was deployed behind the 20-cm aperture Coronal One Shot (COS) coronagraph \citep{Smartt_Fisher_1981} and is now mounted on the spar at the Mauna Loa Solar Observatory (MLSO). CoMP is comprised of: 1) an occulting disk, located at the focus of the COS, that blocks the light from the solar disk; 2) a lens that collimates the solar image; 3) a filter wheel holding three order-blocking filters corresponding to each of the three observable emission line regions; 4) the polarimeter/tunable filter package; 5) a re-imaging lens that forms the final solar image; and 6) a 1024x1024 pixel HgCdTe infrared detector array. The CoMP filter is a four-stage, wide-field calcite birefringent filter with a bandwidth of 1.3 \AA{} (an instrumental width of 21 km/s) and is tuned in wavelength by four liquid crystal variable retarders and has a full field of view of 2.8 R$_{\sun}$ at a spatial sampling of 4.5\arcsec. The data studied in this paper are the ``Dynamics 3" data which take three wavelength positions at the 10747 \AA{} \ion{Fe}{13} line and cadence of 30 s. All of the data analyzed below are openly available on the CoMP webpage (\url{http://www.cosmo.ucar.edu/COMP.html}). 

The reduced CoMP FITS data contain four components: line peak intensity, Doppler velocity, line width and the enhanced intensity. We have found forty-six isolated (without significant line-of-sight complexity) bright coronal loops sets for study that can be grossly grouped as belonging to coronal cavities, active regions and trans-equatorial systems. We then compared these loops observed in the CoMP enhanced line peak intensity images with those tracked through a ``wave-tracking" method, which employs the wave-propagation-angle map generated by a cross-correlation method on the Doppler velocity images \citep{McIntosh_2008}. Nine of these loops have been discarded due to unsatisfactory results from the wave-tracking method (errors in phase speeds exceeding $100 \,\rm{km\,s}^{-1}$ ), which may be a result of line-of-sight superposition effects. For the remaining loops we use CoMP data with less than three missing frames with at least 90 minutes of continuous observation. Any data gaps are filled by linear interpolation using the preceding and following images. As in \cite{Threlfall_2013}, we select six points along a given coronal loop to define an arc using spline interpolation (see for example Fig. \ref{L1OV}B). The coordinates of the arc are subsequently resampled to be equally spaced, where the spacing is chosen to be the CoMP pixel size of 4.5\arcsec (3.24 Mm) for simplicity. For every position along the loop, data are also sampled along a $\sim$20 Mm perpendicular cut, again spaced by the 4.5\arcsec{} (3.24 Mm) CoMP resolution and hence building up a grid of perpendicular cuts centred on the arc.

\section{Example: The Cavity Loops of Sept. 22$^{nd}$ 2013}\label{sect:cavity}
In this section, we present a detailed analysis of a (long) coronal loop that was part of a coronal cavity on Sept. 22$^{nd}$ 2013 as an example \citep[cf.,][]{DeMoortel_2014}, and then compare the results with those from two shorter loops in the same cavity.

\begin{figure*}[tbh]
\begin{center}
\includegraphics[width=\hsize]{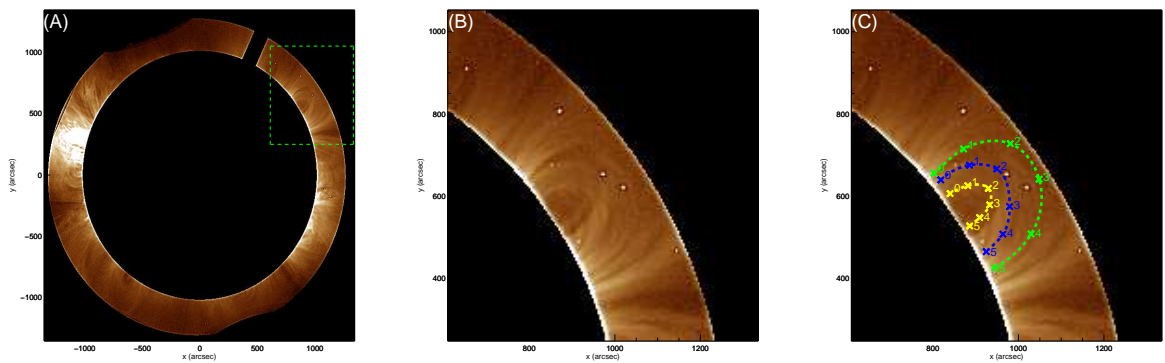}
\caption{(A) Full FOV observation of CoMP enhanced line peak intensity on Sept. 22$^{nd}$ 2013. (B) and (C)  Zoom-in image of the region enclosed in the green box in panel (A). The colored dashed arcs in panel (C) are chosen along the loops, with the green one the long loop, the blue one the medium loop and the yellow one the short loop.}\label{L1OV}
\end{center}
\end{figure*}

\begin{figure*}[t!]
\begin{center}
\includegraphics[width=\hsize]{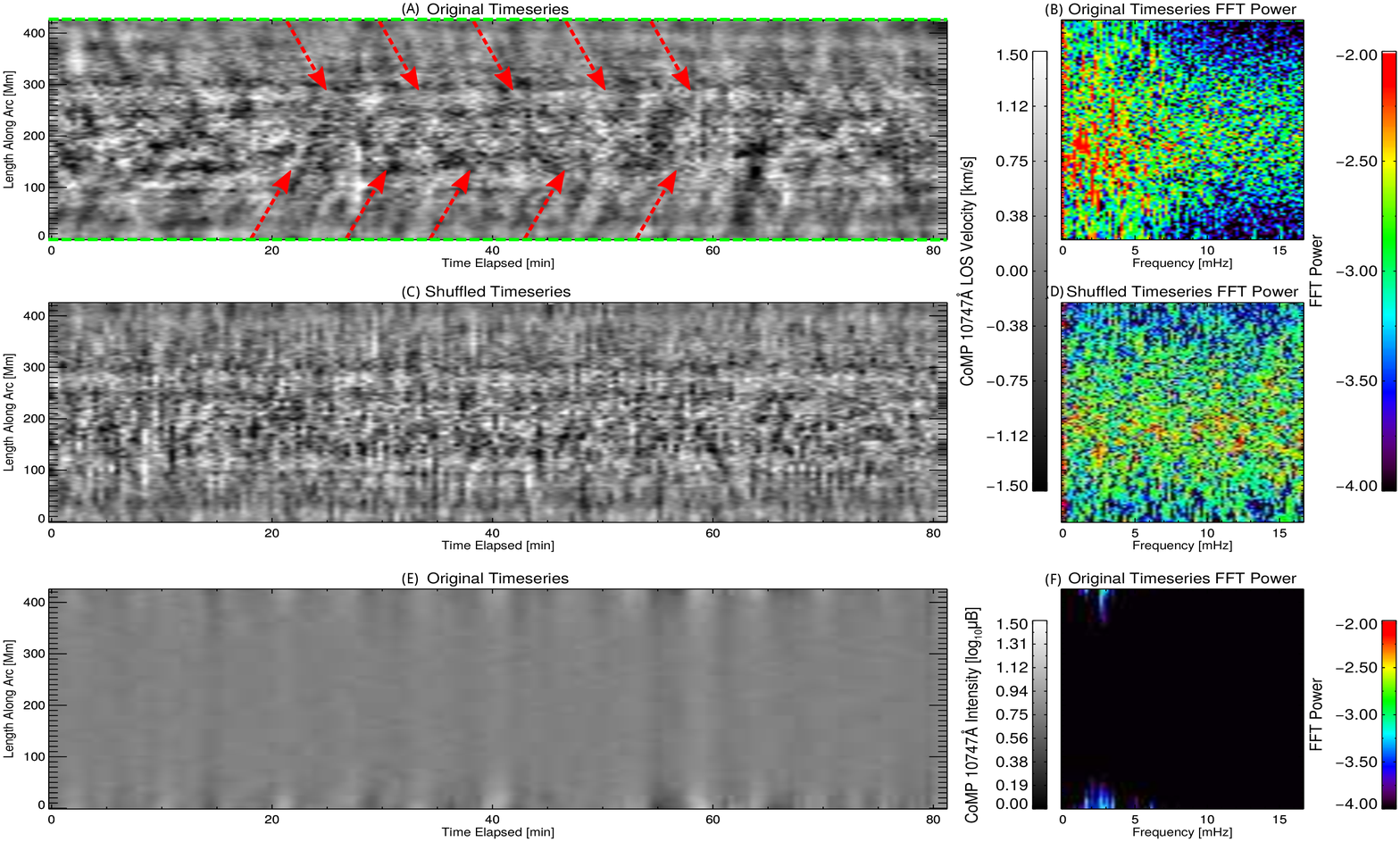}
\caption{(A) and (C): Time-distance plots of Doppler velocity along the arc highlighted by the green curve in Fig. \ref{L1OV}C, based on the original and reshuffled time series, respectively. Perturbations with phase speed around 640 km/s are indicated by red dashed arrows in panel (A). Two green dashed horizontal lines represent the footpoints of the arc. (B) and (D): Corresponding FFT power spectra of (A) and (C) as functions of distance and frequency. FFT power in panel (B) for the original time series and (D) for the reshuffled time series. (E) and (F): Corresponding time-distance plot of line peak intensity along the arc based on the original time series and its associated FFT power spectrum.}\label{L1TD}
\end{center}
\end{figure*}

Fig.~\ref{L1OV}A shows the CoMP enhanced line peak intensity for the full FOV on Sept. 22$^{nd}$ 2013. The green box outlines the region of the cavity on the north-west limb shown where Figs.~\ref{L1OV}B and C show a close-up view of the cavity region. In this region we have identified three clearly isolated loops, outlined in green (the long loop), blue (the medium loop) and yellow (the short loop), respectively, and with position angle (PA) of 57$^\circ$, where PA is measured in degrees from solar North to the apex of the loop. 

Let us start with the analysis of the long (green) loop. As described in Section \ref{sect:instr}, we repeat the \textbf{spline-fitting} procedure of \cite{Threlfall_2013} by choosing 6 spline points along the loop, shown as six green stars labeled from 0 (footpoint 1) to 5 (footpoint 2) in Fig.~\ref{L1OV}C. These spline points are used to generate an arc, shown in green in Fig. \ref{L1OV}C, with the spacing along the arc equal to the CoMP resolution of 4.5\arcsec (3.24 Mm). By integrating the distance between pixels along the arc, the length of the arc is calculated to be $\sim$305\arcsec ($\sim$420 Mm). We should point out, however, that the length of the arc does not equal the real (physical) length of the loop due to (i) projection effects and (ii) the fact that the spline points (0) and (5) are not located at the exact footpoints of the loop. To compensate for the latter effect, we add the distance from the footpoints to the solar limb to the length of the arc, giving a total length of 452 Mm. Projection effects have not been taken into account as it is difficult to exactly identify the corresponding object in {\em STEREO} images for a loop found at the limb of CoMP.

Fig.~\ref{L1TD}A shows the time-distance plot of the CoMP Doppler velocity, along the arc highlighted in Fig.~\ref{L1OV}C by the dashed green line. A clear ``herringbone'' pattern of perturbations (red dashed arrows in Fig.~\ref{L1TD}A), originating from both of the loop footpoints (green dashed lines in Fig.~2A), is visible in the time-distance diagram (similar to the example analyzed by \citealt{DeMoortel_2014}). These perturbations are not generally seen to travel all the way along the loop as they rarely appear to reach the opposite footpoint \-- a tantalizing signature that the waves are changing in passage. The time-distance pattern is less distinct near the loop apex, possibly due to the interaction of the perturbations traveling upwards from the opposite footpoint.

The phase speed of the perturbations is computed using the cross-correlation method developed by \cite{Tomczyk_McIntosh_2009}. We cross-correlate the time series at each position along the arc with the timeseries at the midpoint of the arc (i.e.~the apex of the loop). The peak of the cross-correlation function is then fitted with a parabola such that lag or lead time at each point along the arc is returned. We then fit the lag/lead times versus the distance along the loop with a straight line - the phase speed (and the associated error in the phase speed) of the propagating perturbations are the gradient of this line. When waves are counter propagating this technique gives rise to anomalously high phase speeds (see \citealt{Tomczyk_McIntosh_2009} for more details). Therefore, to obtain more accurate phase speeds we use the technique of \cite{Tomczyk_McIntosh_2009} to identify the phase speeds of waves moving in either direction along the loop: pro/retro-grade filtering is done by masking the positive/negative frequency halves of the $k-\omega$ diagram generated from the FFT of the original time-distance plot, and then performing the inverse transform to construct two space time plots, one for each direction of propagation. The cross-correlation phase speed method is applied to each to yield the (mean) phase speed of the wave on that arc, where the two propagation directions mostly show very similar phase speeds.

By averaging the phase speeds obtained from the filtered time-distance plots employing the cross-correlation method described above, the phase speed of perturbations (waves) propagating along the loop is estimated at 640 ($\pm 34$) km/s (red dashed arrows in Fig.~\ref{L1TD}A). It is substantially larger than the (estimated) local sound speed of a cavity \citep[$\sim$100\ km/s,][]{Liu_2012}. As in previous studies \citep[e.g.,][]{Tomczyk_2007, Tomczyk_McIntosh_2009,Threlfall_2013, DeMoortel_2014} we found little evidence of  simultaneous intensity variations (Fig.~\ref{L1TD}E), indicating that the traveling perturbations are largely non-compressive ({\Aic}) in nature.

Fig.~\ref{L1TD}B shows the logarithm of the FFT power of the time-distance data in Fig.~\ref{L1TD}A as a function of frequency and distance along the arc shown in Fig. \ref{L1OV}. Detailed investigation of the FFT power diagram reveals that: (1) most of the power is concentrated at the lower frequency range, as shown with red colors in the left portion of the diagram; (2) the FFT power at the apex (middle part along the y-axis direction) is consistently higher than at the footpoints (edge regions with maximum and  minimum y value) for all frequencies. This higher power could be caused by linear superposition of the perturbations propagating upwards from both footpoints and/or the effect of gravitational stratification, which leads to the growth of velocity amplitudes with height \citep[see e.g.,][]{Wright_Garman_1998}. As a baseline comparison, we perform a random reshuffle of the original time series \citep{DeMoortel_2014}. The corresponding time-distance plot and FFT power spectrum are shown in Fig.~\ref{L1TD}C and~D. It is clear that there is no longer any evidence of the characteristic herringbone pattern of perturbations. The corresponding FFT power spectrum is almost uniform \-- power is equally distributed over all frequencies. 

\begin{figure*}[tbh!]
\begin{center}
\includegraphics[width=0.8\hsize]{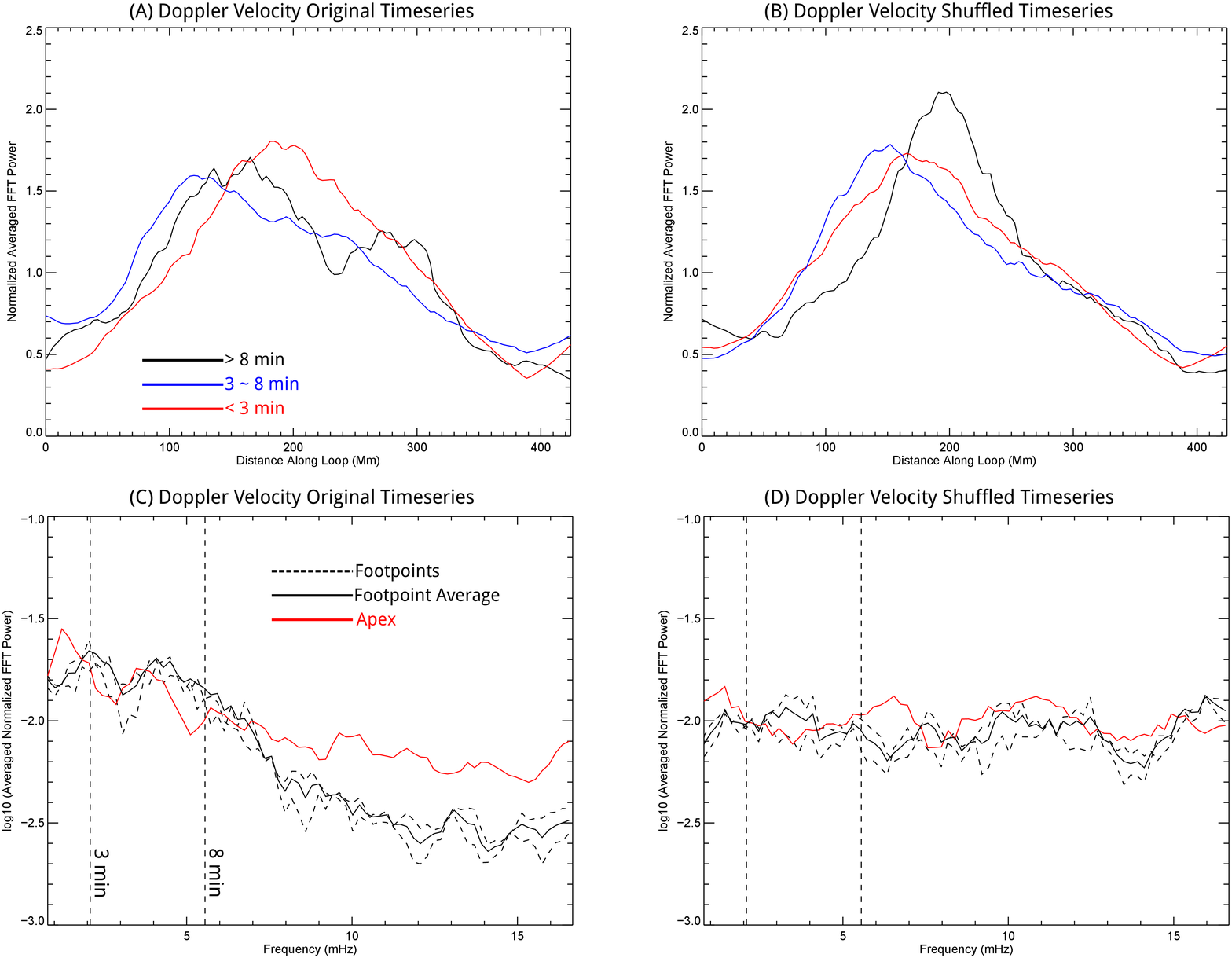}
\caption{(A) and (B): The normalized averaged FFT power at low ($<$ 3 min, black curve), medium (between 3 and 8 min, blue curve) and high ($>$8 min, red curve) frequency ranges (referred to as LF, MF and HF respectively) from the original and reshuffled time series data of the Doppler velocity observation by CoMP on Sept. 22$^{nd}$ 2013, respectively. (C) and (D): The logarithm of the averaged normalized FFT power at two footpoints (black dashed curves) and the apex (red solid curves) as functions of frequency, based on the original and reshuffled time series data of Doppler velocity, respectively. The black solid curve is the average of the FFT power at two footpoints. The two vertical dashed line represent periods of 3 min and 8 min, respectively.}\label{L1FT}
\end{center}
\end{figure*}

To investigate whether the superposition and/or gravitational stratification effects could fully account for the growth of the FFT power at the apex, we average the FFT power across the low ($>$ 8 min), medium (between 3 and 8 min), and high ($<$ 3 min) frequency ranges (hereafter referred to as LF, MF and HF, respectively). The resulting averaged power for the three frequency ranges as a function of distance along the arc is shown in Fig.~\ref{L1FT}A. The curves have been smoothed to remove the smallest scale variations and have been normalized to their own averages to allow easy comparison. It is clear from Fig. \ref{L1FT}A that the FFT power for the three frequency ranges grows at the apex. However, the growth rate for these three curves appears to differ - the FFT power of the higher frequencies grows faster than that of the lower frequencies. Following the estimate of \cite{DeMoortel_2014}, the wave amplitude could increase by a maximum factor of 3.4 at the loop apex, using linear superposition (a maximum factor of 2) and an $e^{z_{apex}/(4H)} \approx 1.7$ gravitational growth  \citep[see e.g.,][]{Wright_Garman_1998} with the gravitational scale height $H=75$ Mm and the height of the loop apex $z_{apex}=160$ Mm for this particular loop.  Fig.~\ref{L1FT}A shows that the FFT power for the LF and MF parts is about 1.36 higher at the apex than at foot points, indicating a $\sqrt{1.36}\approx 1.17$ growth in wave amplitude (as the FFT power scales as the square of the amplitude), which corresponds to a damping of up to 65 \% compared to the ideal non-damping estimation (as 1.17 is about 35\% of 3.4, the maximum possible growth rate). On the other hand, the damping of the HF part is estimated to be roughly 55\%, less than that of the MF and LF parts, implying the presence of an additional effect other than the linear superposition and gravitational stratification. The results for the randomly shuffled time series (Fig.~\ref{L1FT}B) show similar growth rates for the three frequency ranges.

Fig.~\ref{L1FT}C shows the logarithm of the averaged FFT power over distance as a function of frequency. To remove the influence of linear superposition and gravitational stratification effects, the FFT power at each position along the loop is divided by the total FFT power over the whole frequency range at that position. The two dashed black curves show the FFT power at the two footpoint regions (the first and last 20 points of the arc) and the solid curve is their average. The solid red curve is the FFT power around the apex (the middle 20 points of the arc). The FFT power decreases with frequency at the footpoints as well as at the apex. However, a detailed comparison of the two solid curves reveals that the power at the apex (red curve) is less than or at most equal to that of the footpoints (black curve) in the LF and MF ranges, but becomes higher in the HF range. The ratio of the power at the apex and footpoints  is about 1.01 and 0.80, in the LF and MF frequency ranges, respectively. However, the ratio grows to about 1.81 in the HF range, implying an `excess' of  HF power. The randomly shuffled time series on the other hand (Fig. \ref{L1FT}D) shows an even distribution of the FFT power as a function of frequency for the three different regions. To quantify the excess HF FFT power, we define a variable ``Ratio Difference" (RD) which represents how much the ratio (R) between the power at the apex and footpoints grows at the HF range with respect to the LF and MF ranges, i.e. 
 
 $$RD=\frac{R_{HF}-(R_{LF}+R_{MF})/2}{(R_{LF}+R_{MF})/2}\times 100\%$$
 
 \noindent where $R_{LF}$, $R_{MF}$ and $R_{HF}$ are the ratios between the power at the apex and footpoints in the low, medium and high frequency ranges, respectively. Excess high frequency power at the loop apex is characterized by positive values of the Ratio Difference (the high frequency component power decreases slower than the low frequency and medium frequency components when traveling along the loop from the footpoints to the apex) whereas a negative value of the Ratio Difference would indicate that the high frequency component decreases faster along the loop than the low/medium frequency power. The Ratio Difference (RD) value for this particular loop is about 101.56\%, indicating a significant excess of high frequency power at the loop apex.

\begin{figure*}[t!]
\begin{center}
\includegraphics[width=0.8\hsize]{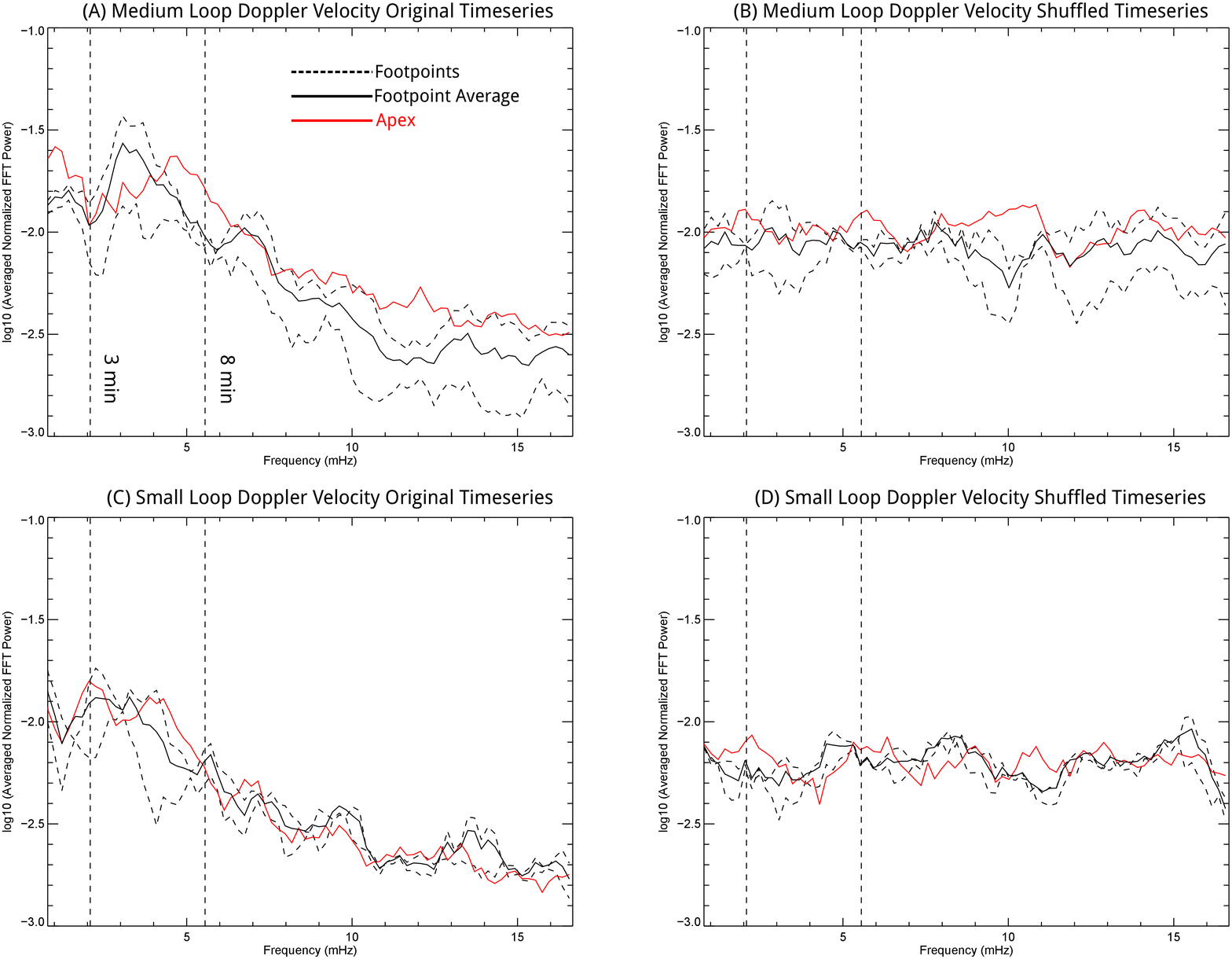}
\caption{(A) and (B): The logarithm of the averaged, normalized FFT power at the two footpoints (black dashed curves) and the apex (red solid curves) as a function of frequency for the medium loop, based on the original and reshuffled time series data of Doppler velocity. The black solid curve is the average of the FFT power at the two footpoints. (C) and (D): same result as panels (A) and (B) but for the short loop. The two vertical dashed lines represent periods of 3 min and 8 min, respectively. }\label{L1OT}
\end{center}
\end{figure*}

In addition to the long (green) loop, we have also analyzed the two shorter loops, outlined by the blue and yellow curves in Fig.~\ref{L1OV}C. We found similar perturbations propagating upwards along the loops, with typical herringbone patterns in the corresponding Doppler velocity time-distance diagrams (not shown). However, the behavior of the EHFF in these two loops is of particular interest. Fig.~\ref{L1OT} shows the corresponding FFT power spectra (averaged and normalized as before) as a function of frequency for these two loops. The FFT power spectrum of the medium loop shows similar behavior as for the long loop:  {the low-frequency FFT power at the apex is lower than the corresponding LF power at the footpoints, but at high frequencies, the power at the apex is higher than at the footpoints (i.e.~ in Fig.~\ref{L1OT}A the red line (apex) falls below the solid black line (footpoints) at low frequencies but above the black line at high frequencies). } The effect is not as pronounced though as for the long loop described earlier, with an RD value of about 17.92\%. However, the EHFF appears to be absent for the shortest loop (Fig.~\ref{L1OT}C). The FFT power at the apex is now approximately equal to the FFT power at the footpoints for all frequencies, leading to an RD value of -16.06\%. This intriguing result of different behavior of the HF power in different loops implies that the EHFF phenomenon is not necessarily present in all loops. All three loops studied in this Section are located in the same cavity and hence it is likely that they share some properties such as their magnetic field topology (and possibly their magnetic field strength and plasma density). The most clear distinction between them is their lengths, namely 452 Mm, 304 Mm and 201 Mm, respectively.

\section{Statistical Analysis of 37 Coronal Loops}\label{sect:stats}

\begin{figure*}[tbh]
\begin{center}
 \includegraphics[width=0.8\hsize]{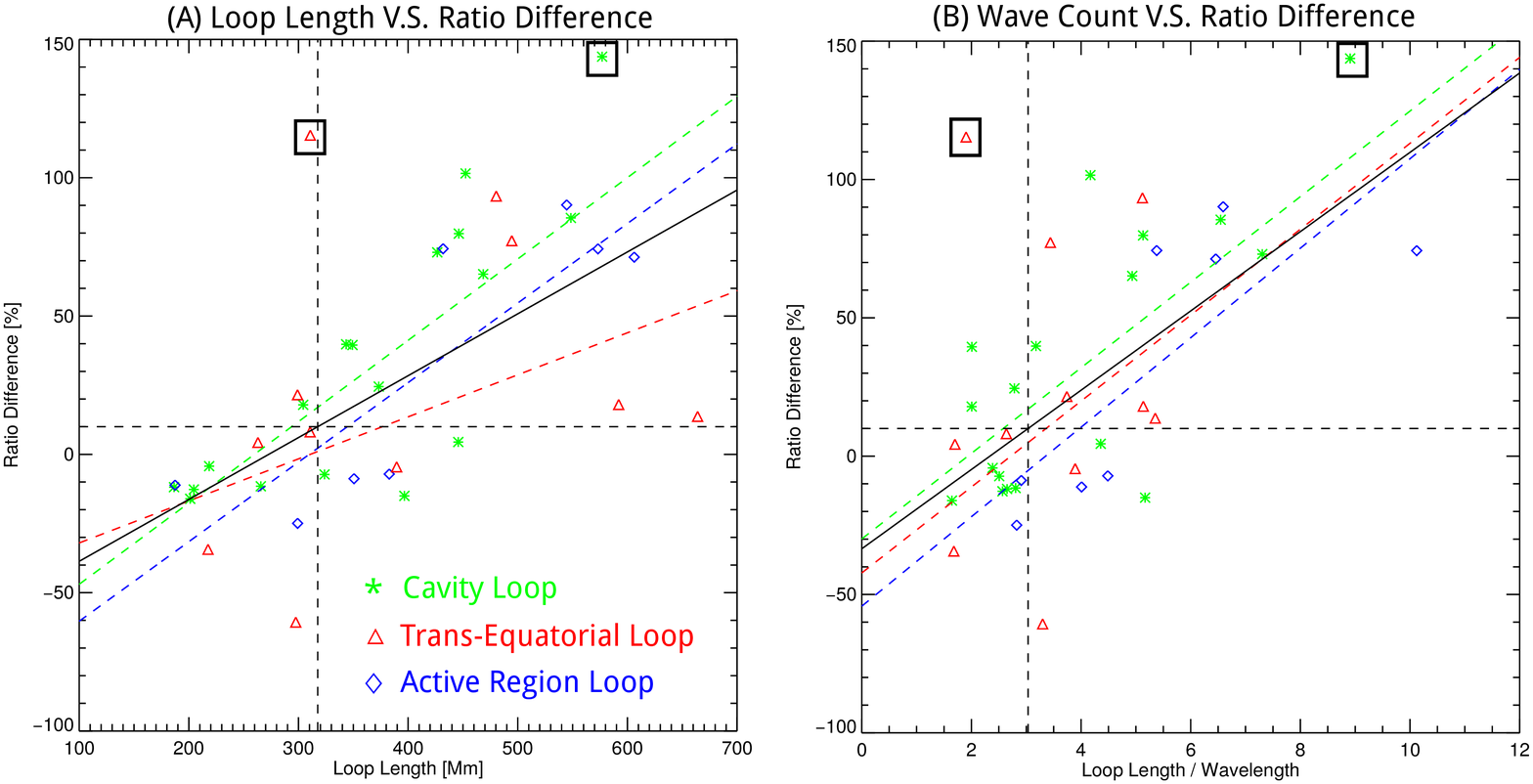}
 \caption{(A) Relationship between the loop length and the Ratio Difference (RD). (B) Relationship between wave count and Ratio Difference. Green asterisks: cavity loops. Blue diamonds: active region loops. Red triangles: trans-equatorial loops. The colored dashed lines are calculated from corresponding linear fits. The black solid line is the linear fit result for the full dataset. The points in black rectangles are found to strongly bias the linear fits and hence they are excluded from further analysis.}\label{WDR}
 \end{center}
\end{figure*}

In the previous section we presented the detailed analysis and the comparison of three cavity loops on Sept. 22$^{nd}$ 2013. However, since those loops are in the same structure, they may not have much difference in magnetic field topology and strength and plasma density. We have seen that the {\Aic} perturbations propagating along those loops reveal quite different behavior \-- the EHFF phenomenon can be easily found in the long and medium loops, but not in the short loop. As noted above, we suppose that the value of RD (representing the significance of the EHFF phenomenon in a loop) may be proportional to the loop length. To examine our supposition, we have analyzed thirty-four more coronal loops using the same methods presented previously. As noted earlier, these loops are chosen if they are clearly (and easily) isolated in the CoMP FOV.

Table \ref{tb} provides the measured parameters for these loops: ``YMD" is the date of the observation; ``PA" is the position angle of the apex of a loop with respect to the due north; ``L" is the apparent length of the loop in the plane of the sky; ``RD" is the ``Ratio Difference" defined in Sect.~3; ``$V_{phase}$" is the {\Aic} perturbation phase speed detected in a loop using the cross-correlation method described in Sect.~3; and ``WD" is the line width difference and represents how much the average line width increases or decreases at the apex with respect to the footpoints.

The parameter $\lambda$ represents the ``characteristic wavelength'' of the {\Aic} perturbations propagating in a loop and is defined as follows. We integrate the Doppler velocity FFT power over three narrow wavelength ranges: 1.5$\pm$0.3~mHz, 3.5$\pm$0.5~mHz and 7.0$\pm$1.0~mHz. We then fit the relationship between the three integrated FFT powers and the filter frequencies with a Gaussian \-- the characteristic frequency is that for which the fitted Gaussian function peaks. The quotient of the wave phase speed and the characteristic frequency we then define as the characteristic wavelength for the loop.

Based on the limited sample in Sect. 3 we speculated that the value of RD ratio varies with loop length. As the longest loop shows a significant excess of high frequency FFT power and the shortest loop reveals evidence of roughly equal or even higher damping at the HF part \-- we believe that the RD metric represents the degree by which the EHFF phenomenon increases with loop length. Plotting the tabulated results in Fig.~\ref{WDR}A would appear to add weight to our early speculation as there appears to be a linear relationship between loop length and the Ratio Difference.

Fig.~\ref{WDR} allows us to analyse different types of coronal loops prevalent in CoMP data: cavity loops, active region loops and trans-equatorial loops. They are represented by green asterisks, blue diamonds and red triangles, respectively. It is clear that the RD value grows with the loop length and the relationship between them appears to be linear at least within the loop length range we study of 200 Mm to 700 Mm and RD range of -61\% to 144\%. The slopes of the green, blue and red dashed lines are 0.295, 0.287 and 0.152, respectively. The solid black line represents the linear fit result of the whole dataset with a relatively high cross-correlation factor (CC) of about 0.7, implying a reliable proportional relationship between the loop length and RD. We note that the linear fits can be strongly biased by the two data points enclosed in rectangles \-- it is not immediately clear what is incorrect or wrong with these measurements. but we exclude them from further analysis. Using a Ratio Difference of 10\% as a threshold value to indicate excess high frequency power, the overall linear fit (black line) allows us to determine a critical loop length: when a loop is longer than 318 Mm, the {\Aic} perturbations appear to damp slower at high frequencies when propagating from the footpoints to the apex than they do at low frequencies. The loop studied by \cite{DeMoortel_2014} has a length of $\sim480\ Mm$ and do indeed exhibit this behavior.

{\Aic} perturbations propagating along different loops could display different behavior, considering that loops may have different properties \citep[e.g. density, magnetic field strength,É][]{Priest_1978}.  For the propagating waves considered in this study, wavelength is one representation for plasma density, magnetic field strength and wave frequency. As shown in Fig.~\ref{WDR}B, RD has a proportional relationship with the ratio of the loop length and the characteristic wavelength, a measure we define as the ``wave count'' \-- or the number of waves wholly contained in the loop. Again, we fit the relationship with a linear function and the cross-correlation factor turns out to be larger than 0.5 ($\sim$0.6). Following the definition of critical length, we find the value of the critical wave count to be about 3.0$\pm$0.9 for RD to be large enough (shown as the black dashed lines in Fig.~\ref{WDR}B). 

\section{Discussion}\label{sect:disc}
We analyzed thirty-seven clearly isolated CoMP coronal loops. In all cases, {\Aic} perturbations are found propagating along coronal loops. Detailed analysis reveals that these perturbations cause obvious variations in Doppler velocity but not in simultaneous intensity data, implying their incompressible ({\Aic}) nature, as reported in \cite{Tomczyk_2007}. The phase speed of these perturbations ranges from 250 km/s to 750 km/s and their typical period varies from 140 s to 270 s (roughly 3 \--5 min) \-- consistent with those reported in \cite{Tomczyk_2007} and \cite{Tomczyk_McIntosh_2009}.

Further, time-distance analysis of the Doppler signal in a coronal loop of Sept. 22$^{nd}$ 2013 ($\sim$450 Mm long) revealed a herringbone pattern, indicating perturbations propagating from both footpoints to the apex, similar to \cite{Tomczyk_McIntosh_2009}. The bi-directional perturbations have almost the same phase speed and they interact with each other around the apex. No obvious downward propagating perturbations around the footpoints are found, implying that these perturbations damp continuously on approach to, and after passing, the loop apex.

\begin{center}
\begin{longtable}{c|c|c|c|c|c|c|c}
\multicolumn{7}{r}%
{\tablename\ \thetable\ : \textit{Statistical results of thirty-seven coronal loops in 2013}} \\ 
\hline \label{tb} 
YMD & Type & PA ($^\circ$) & L (Mm) & RD (\%) & V$_{phase}$ (km s$^{-1}$) & $\lambda$ (Mm) & WD (\%) \\  
\hline 
\endfirsthead 
\multicolumn{6}{r}%
{\tablename\ \thetable\ : \textit{Continued From Previous Page}} \\ 
\hline 
YMD & Type & PA ($^\circ$) & L (Mm) & RD (\%) & V$_{phase}$ (km s$^{-1}$) & $\lambda$ (Mm) & WD (\%) \\  
\hline 
\endhead 
\hline \multicolumn{6}{r}{\textit{Continued On Next Page}} \\
\endfoot 
\hline \hline
\endlastfoot
20130104 & CV & 313.47 & 396.56 & -15.01 & 309.45$\pm$19.43 & 76.73 & 3.67\\
20130202 & TE & 90.94 & 299.14 & 21.56 & 332.75$\pm$18.17 & 79.97 & 5.43\\
20130308 & AR & 57.96 & 431.85 & 74.34 & 350.89$\pm$16.62 & 80.27 & 9.90\\
20130308 & AR & 70.02 & 187.41 & -11.13 & 231.86$\pm$34.94 & 46.73 & 16.61\\
20130308 & TE & 98.88 & 480.31 & 93.37 & 386.82$\pm$31.47 & 93.82 & 10.33\\
20130308 & CV & 210.45 & 548.42 & 85.46 & 413.91$\pm$23.99 & 83.78 & 10.64\\
20130308 & AR & 260.21 & 544.61 & 90.17 & 386.86$\pm$33.31 & 82.61 & 26.60\\
20130313 & TE & 97.11 & 297.52 & -60.63 & 631.29$\pm$34.23 & 90.18 & 26.15\\
20130313 & AR & 114.03 & 350.61 & -8.78 & 607.34$\pm$33.26 & 120.50 & 29.63\\
20130403 & TE & 92.13 & 310.59 & 8.13 & 541.18$\pm$41.84 & 117.74 & 27.28\\
20130418 & TE & 267.29 & 664.01 & 13.74 & 571.21$\pm$39.35 & 124.05 & 8.96\\
20130418 & TE & 269.04 & 389.53 & -4.50 & 399.54$\pm$16.20 & 100.07 & 4.12\\
20130502 & TE & 89.29 & 494.53 & 77.24 & 577.09$\pm$23.46 & 143.72 & 17.10\\
20130502 & TE & 91.64 & 263.02 & 4.31 & 608.33$\pm$42.61 & 154.73 & 2.94\\
20130515 & TE & 270.90 & 592.04 & 18.02 & 466.47$\pm$21.55 & 115.31 & 4.32\\
20130623 & AR & 101.67 & 299.20 & -24.93 & 490.29$\pm$29.21 & 105.90 & 17.22\\
20130627 & CV & 159.39 & 343.49 & 39.79 & 467.83$\pm$27.73 & 108.10 & 10.88\\
20130702 & CV & 60.32 & 349.22 & 39.52 & 746.17$\pm$39.99 & 173.94 & 11.48\\
20130708 & AR & 235.10 & 573.15 & 74.29 & 309.02$\pm$22.87 & 56.62 & 20.43\\
20130712 & AR & 88.97 & 382.76 & -7.12 & 435.04$\pm$42.98 & 85.29 & 24.40\\
20130715 & CV & 49.35 & 218.54 & -4.26 & 429.79$\pm$30.48 & 91.55 & 7.53\\
20130914 & CV & 199.60 & 576.91 & 143.75 & 316.02$\pm$32.02 & 64.75 & 15.72\\
20130914 & TE & 93.67 & 310.76 & 115.36 & 740.61$\pm$29.19 & 163.35 & 31.48\\
20130914 & TE & 94.08 & 217.51 & -34.33 & 619.78$\pm$14.96 & 129.48 & 8.39\\
20130921 & CV & 59.20 & 468.50 & 65.15 & 416.70$\pm$28.86 & 94.96 & 8.30\\
20130921 & CV & 60.88 & 323.74 & -7.25 & 537.96$\pm$28.91 & 129.12 & 2.77\\
20130921 & CV & 60.29 & 204.55 & -12.69 & 334.31$\pm$16.10 & 79.69 & 7.21\\
20130922 & CV & 57.52 & 452.39 & 101.56 & 460.82$\pm$28.59 & 108.49 & 6.98\\
20130922 & CV & 57.53 & 304.30 & 17.92 & 621.46$\pm$19.50 & 151.77 & 3.17\\
20130922 & CV & 57.27 & 201.22 & -16.06 & 506.51$\pm$28.18 & 122.37 & 8.10\\
20130923 & CV & 53.71 & 373.20 & 24.50 & 588.99$\pm$24.49 & 134.08 & 7.05\\
20130923 & CV & 54.44 & 186.49 & -11.86 & 304.71$\pm$20.89 & 70.37 & 2.82\\
20131005 & CV & 202.44 & 426.54 & 73.03 & 250.88$\pm$11.94 & 58.39 & 8.68\\
20131005 & CV & 296.66 & 445.73 & 4.47 & 452.36$\pm$52.04 & 102.29 & 4.72\\
20131005 & CV & 323.14 & 265.63 & -11.53 & 366.89$\pm$17.27 & 94.66 & 7.62\\
20131005 & CV & 324.84 & 446.22 & 79.79 & 361.48$\pm$22.64 & 86.95 & 3.85\\
20131102 & AR & 298.75 & 606.16 & 71.29 & 399.28$\pm$15.97 & 93.87 & 3.64\\
\end{longtable}
YMD: Date. Type: Three types of loops, Cavity (CV), Active Region (AR), Trans-Equatorial (TE). PA: Position angle of the loop with respect to the arctic pole in units of degree. L: Loop length in units of megameter. RD: Ratio difference (see text). V$_{phase}$: Propagating perturbation phase speed in units of kilometer per second. $\lambda$: Characteristic wave length, in units of megameter. WD: Line width difference (see text).
\end{center}

FFT power analysis shows that these perturbations undergo damping when propagating from the footpoints to the apex across the frequency range accessible to CoMP. One possible means of damping these waves is through ``mode coupling" \-- a process inherent to transverse (kink) waves propagating in a loop with an inhomogeneous boundary layer and where the (observed) wave damping occurs due to the transfer of the wave energy from the transverse waves generated at the loop footpoint regions into azimuthal {\Aic} waves as they propagate along the loop \citep[e.g.,][]{Melrose_1977, Pascoe_2010, Hood_2013, Pascoe_2013}. 

 The basic mode coupling process has an inherent frequency dependence: high frequency waves damp faster than low frequency waves \citep[e.g.,][]{Pascoe_2010,Terradas_2010}. This means that we would expect high frequency FFT power at the loop apex to be lower than observed. Further, in a mode coupling regime, high frequency perturbations damp faster in higher density loops (that are typically shorter) \citep{Terradas_2010}. This relationship {\em could} explain the different inclinations between the linear fit lines (Fig.~\ref{WDR}A and~B to cavity loops (green dashed line), active region loops (blue dashed line) and trans-equatorial loops (red dashed line). For example, as the plasma density in active and trans-equatorial regions is likely to be larger than cavities, lower inclinations would be expected. However, further investigation is required to confirm this suggestion.


Given the frequency selectivity in mode coupling, less high-frequency power would be expected at the loop apex compared to lower frequencies. However, this is contrary to what the observations and analysis of the long and medium loops show. As shown in Fig.~\ref{L1FT}, the high frequency FFT power at the apex of these perturbations is higher than that at the low- and medium-frequencies. Recalling that we defined the Ratio Difference as a measure of how much the high frequency power is larger than the low/medium frequency power, one would expect the value of RD to be always below 0 if the high-frequency part damps faster (as in the mode coupling model). However, this is not the case and the RD is larger than 0 for the long and medium loops. \cite{DeMoortel_2014} suggested that the excess of the high-frequency power may be evidence of the onset of turbulence \citep[e.g.,][]{vanBallegooijen_2011} caused by the non-linear interaction of two counter-propagating wave trains. Here, counter-propagating wave trains could be either two wave trains traveling upwards from two opposing footpoints, or a wave train interacting with its reflected wave train produced by density variations along the loop. 

Our statistical analysis shows that the Ratio Difference is proportional to the loop length. In other words, the longer the loop is, the more high-frequency power is generated. This seems intuitively to agree with the onset of turbulence due to non-linear interactions between the oppositely traveling wave trains, if a sufficient number of wavelengths is present along the loop.  Fig.~\ref{WDR}B shows the proportional relationship between the wave count (representing the number of wavelengths along the loop) and the Ratio Difference. This plot reveals that the more wave lengths there are along the loop, the more high-frequency power is generated \-- this again would tend to support the premise that turbulence is present.

\begin{figure*}[tbh]
\begin{center}
 \includegraphics[width=0.8\hsize]{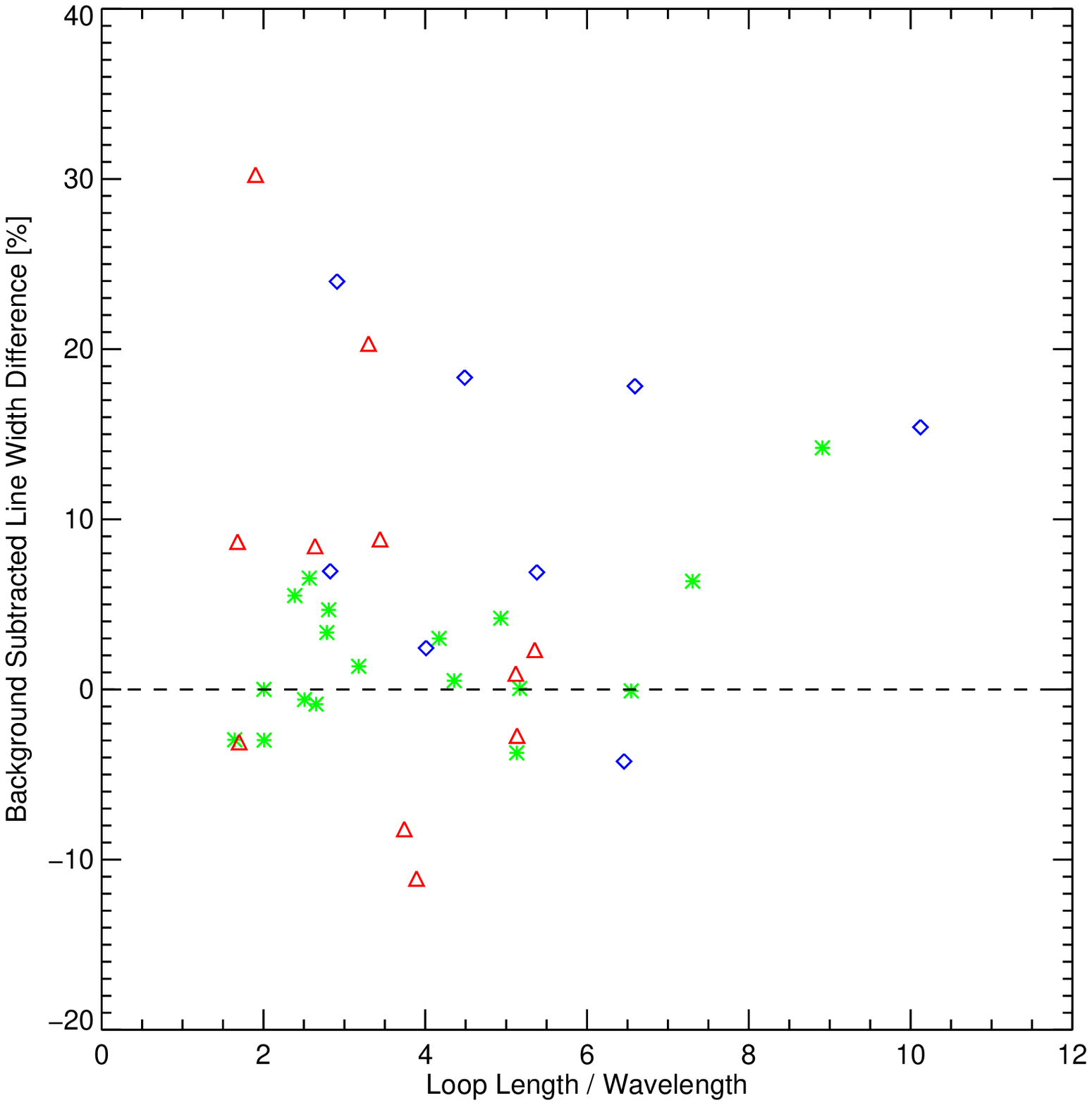}
 \caption{ Relationship between the wave count
 		and the background subtracted line width difference. Green asterisks: cavity loops. Blue diamonds: active region loops. 
		Red triangles: trans-equatorial loops.}\label{WDW}
\end{center}
\end{figure*}

If turbulence is present, the line width of optically thin lines like those observed by CoMP would be expected to increase at the apex. However, many other effects influence line widths. For example, gravitational stratification and line-of-sight superposition can increase the observed line widths \citep{McIntrosh_DePontieu_2012}, whereas damping of waves has the opposite effect. This complexity makes studying the direct relationship between the wave count and the line width difference (WD, defined in Section 4 and shown in Table \ref{tb}) meaningless. To subtract the influence of stratification, damping and some other effects that may change the line width with height, we calculate the difference between the average line width at the height of the apex and the footpoints over the whole time sequence with a 1$^\circ$ angular spacing around the Sun as the background line width change (``WDB"). Fig. \ref{WDW} shows the relationship between the wave count and the difference between WD and WDB. We see that $\sim$70\% of the points lie above 0 (the horizontal dashed line) possibly indicating that the line broadening at the majority of the loop apexes sampled is larger than may be expected via waves propagating through a complex structural superposition. However, such superposition likely will not change the loop lengths or wave counts, possibly indicating that the distribution of the points in Fig.~\ref{WDW} could be a signature of turbulence in coronal loops.

\section{Conclusion}\label{sect:conc}

In this paper, following the work by \cite{DeMoortel_2014}, we performed a detailed analysis of three cavity loops observed on Sept 22$^{nd}$ 2013. Doppler velocity perturbations with phase speeds around 640 km s$^{-1}$ propagating from the loop footpoints are found in the corresponding time-distance diagrams without any simultaneous density variations. An excess of high frequency power (EHFF) is found in the FFT power spectrum of the two long loops and not in that of the shortest loop. This EHFF phenomenon might be tentative evidence for the onset of {\Aic} turbulence. Further statistical analysis on thirty-seven clearly isolated loops shows a relationship between the loop length and the Ratio Difference (a measure of the excess high frequency at the apex), in agreement with the assumption of turbulence, as a longer traveling distance could lead to more (non-linear) interactions between opposite-propagating wave trains. Linear fits reveal a loop length of at least about 318 Mm for the EHFF phenomenon to be present.

The proportional relationship between the wave count (how many wavelengths there are along a loop) and the Ratio Difference tends to support the presence of {\Aic} turbulence. The critical wave count of 3.0 for the excess high frequency (EHFF) to be present is consistent with (the onset of) turbulence. Finally, we have explored the relationship between the wave count and the background subtracted line width. The weakly proportional relationship between them again supports the onset of {\Aic} turbulence in coronal loops.

In this study, we have presented statistical evidence for the onset of {\Aic} turbulence in coronal loops. However, the relatively low spatial resolution and signal-to-noise ratio of CoMP data prevent us from performing a more detailed analysis and finding more direct evidence of turbulence in those loops. Hopefully in the future, more detailed observations using instruments with higher spacial resolution and signal-to-noise, combined with numerical simulations will help improve our understanding of coronal turbulence and any potential impact on coronal heating.

\acknowledgements{Acknowledgements}
JL is a student visitor at HAO. JL is supported by the Chinese Scholarship Council (CSC 201306340034) and also supported by Grants from NSFC  41131065, 41121003, 973 Key Project 2011CB811403 and CAS Key Research Program KZZD-EW-01-4. JL also thanks his advisor in China, Dr. Yuming Wang. NCAR is sponsored by the National Science Foundation. CoMP data can be found at \url{http://mlso.hao.ucar.edu/}. We acknowledge support from NASA contracts NNX08BA99G, NNX11AN98G, NNM12AB40P, NNG09FA40C ({\em IRIS}), and NNM07AA01C ({\em Hinode}). The research leading to these results has also received funding from the European Commission Seventh Framework Programme (FP7/ 2007-2013) under the grant agreement SOLSPANET (project No. 269299, \url{www.solspanet.eu/solspanet}).  

\bibliographystyle{agufull}
\bibliography{mybib}

\end{document}